\documentclass[12pt]{article}
\usepackage{amsfonts,bm}
\usepackage{graphicx}
\baselineskip
\offinterlineskip
\newcommand{\n}{\newline}

\begin{document}

\begin{center}
{\bf Spin Hamilton Operators, Symmetry Breaking, 
Energy Level Crossing and Entanglement}
\end{center}

\begin{center}
{\bf Willi-Hans Steeb, Yorick Hardy and Jacqueline de Greef} \\[2ex]
International School for Scientific Computing, \\
University of Johannesburg, Auckland Park 2006, South Africa, \\
e-mail: {\tt steebwilli@gmail.com}
\end{center}

\strut\hfill

\strut\hfill

{\bf Abstract} We study finite-dimensional product Hilbert spaces,
coupled spin systems, entanglement and energy level crossing.
The Hamilton operators are based on the Pauli group. 
We show that swapping the interacting term can lead from unentangled 
eigenstates to entangled eigenstates and from an energy spectrum with 
energy level crossing to avoided energy level crossing. 

\strut\hfill

\section{Introduction}

Let ${\cal H}_1$, ${\cal H}_2$ be Hilbert spaces and
${\cal H}_1 \otimes {\cal H}_2$ be the tensor product Hilbert space
[1,2]. Quite often a self-adjoint Hamilton operator acting on the
the tensor product Hilbert space ${\cal H}_1 \otimes {\cal H}_2$
can be written as
$$
\hat H = \hat H_1 \otimes I_2 + I_1 \otimes \hat H_2 + \epsilon \hat V
\eqno(1)  
$$
where the self-adjoint Hamilton operator $\hat H_1$ acts in the 
Hilbert space ${\cal H}_1$, the self-adjoint Hamilton operator
$\hat H_2$ acts in the Hilbert space ${\cal H}_2$, $I_1$ is the identity
operator acting in the Hilbert space ${\cal H}_1$ and $I_2$ is the 
identity operator acting in the Hilbert space ${\cal H}_2$.
The self-adjoint operator $\hat V$ acts in the product Hilbert space
and $\epsilon$ is a real parameter. The main task would be
to find the spectrum of $\hat H$.
\newline

In the following we consider the finite-dimensional Hilbert
space ${\cal H}_1={\cal H}_2={\mathbb C}^n$ and 
then $\otimes$ denotes the Kronecker product [3,4,5,6].
Let $I_n$ be the $n \times n$ identity matrix.
We consider the two hermitian Hamilton operators
$$
\hat H = \alpha A \otimes I_n + I_n \otimes \beta B + 
\epsilon (A \otimes B)
\eqno(2)
$$
$$
\hat K = \alpha A \otimes I_n + I_n \otimes \beta B + 
\epsilon (B \otimes A)
\eqno(3)
$$
where $A$, $B$ are nonzero $n \times n$ hermitian matrices and $\alpha$, $\beta$, $\epsilon$
are real parameters with $\epsilon \ge 0$. We assume that $[A,B] \ne 0$.
The vector space of the $n \times n$ matrices over $\mathbb C$ form a Hilbert space
with the scalar product $\langle X,Y\rangle:=\mbox{tr}(XY^*)$. We also assume
that $\langle A,B\rangle=0$, i.e. the nonzero $n \times n$ hermitian matrices $A$ and $B$
are orthogonal to each other. Of particular interest would be the case where $A$ and $B$ are 
elements of a semi-simple Lie algebra. We discuss the eigenvalue problem for the two 
Hamilton operators and its connection with entanglement and energy level
crossings for specific choices of $A$ and $B$. In the following the matrices 
$A$ and $B$ are realized by Pauli spin matrices. The Hamilton operator will be a linear
combination of elements of the Pauli group ${\cal P}_n$. The Pauli 
group [7] is defined by
$$
{\cal P}_n := \{ \, I_2, \, \sigma_x, \, \sigma_y, \, \sigma_z \, \}^{\otimes n} 
\otimes \{ \, \pm 1, \, \pm i \, \} \,.
\eqno(4)
$$
Such two-level and higher level quantum systems and their physical realization have been
studied by many authors (see [8] and references therein).
The thermodynamic behaviour is determined by the partition functions
$$
Z_{\hat H}(\beta) = \mbox{tr}(\exp(-\beta \hat H)), \qquad
Z_{\hat K}(\beta) = \mbox{tr}(\exp(-\beta \hat K)) \,.
$$
Since 
$$
\mbox{tr}(\hat H) = \mbox{tr}(\hat K) = \alpha n \mbox{tr}(A)
+ \beta n \mbox{tr}(B) + \epsilon (\mbox{tr}(A)) 
(\mbox{tr}(B)) 
$$
the sum of the eigenvalues of the operators $\hat H$ and 
$\hat K$ are the same. However, in general, the partition
functions will be different.  
\newline

\section{Commutators, Eigenvalues and Eigenvectors}

Let us first summarize the equations we utilize in the following. Let $A$, $B$ be 
$n \times n$ matrices over $\mathbb C$. First note that we have the following commutators
$$
[A \otimes I_n,I_n \otimes B] = 0, \quad
[A \otimes I_n,A \otimes B] = 0, \quad
[I_n \otimes B,A \otimes B] = 0
$$
and 
$$  
[A \otimes I_n,B \otimes A] = ([A,B]) \otimes A, \quad
[I_n \otimes B,B \otimes A] = B \otimes ([B,A]) \,.
$$
The last two commutators would be 0 if $[A,B]=0$.
There is an $n^2 \times n^2$ permutation matrix $P$ (swap gate) 
such that
$$
P(A \otimes B)P^{-1} = B \otimes A \,.
$$
This implies that $P(A \otimes I_n)P^{-1}= I_n \otimes A$
and $P(I_n \otimes B)P^{-1}=B \otimes I_n$.
\newline      

Now let $A$ and $B$ be $n \times n$ hermitian matrices.
If the eigenvalues and normalized eigenvectors of $A$ and $B$ are 
$\lambda_j, {\bf u}_j$, $\mu_j, {\bf v}_j$, $(j=1,2,\dots,n)$,
respectively, then the eigenvalues and normalized eigenvectors of 
the Hamilton operator (2) are given by $[3,4,5,6]$
$$
\alpha \lambda_j + \beta \mu_k + \epsilon \lambda_j \mu_k, \quad 
{\bf u}_j \otimes {\bf v}_k \qquad j,k=1,2,\dots,n \,.
$$
Thus the eigenvectors are not entangled since they can
be written as product states. These results can be extended to the
Hamilton operator
$$
\hat H = \alpha (A \otimes I_n \otimes I_n) + \beta (I_n \otimes B \otimes I_n)
+ \gamma (I_n \otimes I_n \otimes C) + \epsilon (A \otimes B \otimes C) 
$$ 
and higher dimensions.

\section{Pauli Spin Matrices and Entanglement}

Since we realize the linear operators $A$ and $B$ by Pauli spin matrices
we summarize some results for the Pauli spin matrices and their Kronecker
products. Consider the Pauli spin matrices $\sigma_z$, $\sigma_x$, $\sigma_y$. 
The eigenvalues are given by $+1$ and $-1$ with the corresponding
normalized eigenvectors
$$
\pmatrix { 1 \cr 0 }, \,\,\, \pmatrix { 0 \cr 1 }, \qquad
\frac1{\sqrt2} \pmatrix { 1 \cr 1 }, \,\,\, 
\frac1{\sqrt2} \pmatrix { 1 \cr -1 }, \qquad
\frac1{\sqrt2} \pmatrix { -i \cr 1 }, \,\,\,
\frac1{\sqrt2} \pmatrix { i \cr 1 }
$$
for $\sigma_z$, $\sigma_x$ and $\sigma_y$, respectively.
Consider now the three hermitian and unitary $4 \times 4$ matrices
$\sigma_x \otimes \sigma_x$, $\sigma_y \otimes \sigma_y$,  
$\sigma_z \otimes \sigma_z$.
These matrices appear in Mermin's magic square [9] for the proof of 
the Bell-Kochen-Specker theorem. Since the eigenvalues of the Pauli matrices 
are given by $+1$ and $-1$, the eigenvalues of the $4 \times 4$ matrices
$\sigma_x \otimes \sigma_x$, $\sigma_y \otimes \sigma_y$,
$\sigma_z \otimes \sigma_z$ are $+1$ (twice) and
$-1$ (twice). The eigenvectors can be given as product states (unentangled states), 
but also as entangled states (i.e. they cannot be written as product states). Obviously 
$$
\pmatrix { 1 \cr 0 } \otimes \pmatrix { 1 \cr 0 }, \quad
\pmatrix { 1 \cr 0 } \otimes \pmatrix { 0 \cr 1 }, \quad
\pmatrix { 0 \cr 1 } \otimes \pmatrix { 1 \cr 0 }, \quad
\pmatrix { 0 \cr 1 } \otimes \pmatrix { 0 \cr 1 }
$$
are four normalized product eigenstates of $\sigma_z \otimes \sigma_z$.
The normalized product eigenstates of $\sigma_x \otimes \sigma_x$ are 
$$
\frac12 \pmatrix { 1 \cr 1 } \otimes \pmatrix { 1 \cr 1 }, \quad
\frac12 \pmatrix { 1 \cr 1 } \otimes \pmatrix { 1 \cr -1 }, \quad
\frac12 \pmatrix { 1 \cr -1 } \otimes \pmatrix { 1 \cr 1 }, \quad
\frac12 \pmatrix { 1 \cr -1 } \otimes \pmatrix { 1 \cr -1 } \,.
$$
The normalized product eigenstates of $\sigma_y \otimes \sigma_y$ 
are 
$$
\frac12 \pmatrix { i \cr 1 } \otimes \pmatrix { i \cr 1 }, \quad
\frac12 \pmatrix { i \cr 1 } \otimes \pmatrix { -i \cr 1 }, \quad
\frac12 \pmatrix { -i \cr 1 } \otimes \pmatrix { i \cr 1 }, \quad
\frac12 \pmatrix { -i \cr 1 } \otimes \pmatrix { -i \cr 1 } \,.
$$
All three $4 \times 4$ matrices also admit the Bell basis 
$$
\frac1{\sqrt2} \pmatrix { 1 \cr 0 \cr 0 \cr 1 }, \quad
\frac1{\sqrt2} \pmatrix { 0 \cr 1 \cr 1 \cr 0 }, \quad
\frac1{\sqrt2} \pmatrix { 1 \cr 0 \cr 0 \cr -1 }, \quad
\frac1{\sqrt2} \pmatrix { 0 \cr 1 \cr -1 \cr 0 }
$$ 
as normalized eigenvectors which are maximally entangled.
As measure for entanglement the tangle $[5,7,10,11]$ will be utilized.
\newline

Consider now the hermitian and unitary $4 \times 4$ matrices
$\sigma_x \otimes \sigma_z$, $\sigma_z \otimes \sigma_x$.
Since the eigenvalues of the Pauli matrices are given by $+1$ and $-1$,
the eigenvalues of the $4 \times 4$ matrices $\sigma_x \otimes \sigma_z$ and 
$\sigma_z \otimes \sigma_x$, are $+1$ (twice) and $-1$ (twice).
The eigenvectors can be given as product states (unentangled states), 
but also as entangled states (i.e. they cannot be written as product states).
The normalized product eigenstates of $\sigma_x \otimes \sigma_z$ are 
$$
\frac1{\sqrt2} \pmatrix { 1 \cr 1 } \otimes \pmatrix { 1 \cr 0 }, \quad
\frac1{\sqrt2} \pmatrix { 1 \cr 1 } \otimes \pmatrix { 0 \cr 1 }, \quad
\frac1{\sqrt2} \pmatrix { 1 \cr -1 } \otimes \pmatrix { 1 \cr 0 }, \quad
\frac1{\sqrt2} \pmatrix { 1 \cr -1 } \otimes \pmatrix { 0 \cr 1 } \,.
$$
The normalized product eigenstates of $\sigma_z \otimes \sigma_x$ are 
$$
\frac1{\sqrt2} \pmatrix { 1 \cr 0 } \otimes \pmatrix { 1 \cr 1 }, \quad
\frac1{\sqrt2} \pmatrix { 1 \cr 0 } \otimes \pmatrix { 1 \cr -1 }, \quad
\frac1{\sqrt2} \pmatrix { 0 \cr 1 } \otimes \pmatrix { 1 \cr 1 }, \quad
\frac1{\sqrt2} \pmatrix { 0 \cr 1 } \otimes \pmatrix { 1 \cr -1 } \,.
$$
The two $4 \times 4$ matrices also admit 
$$
\frac12 \pmatrix { -1 \cr -1 \cr -1 \cr 1 }, \qquad
\frac12 \pmatrix { -1 \cr 1 \cr 1 \cr 1 }, \qquad
\frac12 \pmatrix { 1 \cr -1 \cr 1 \cr 1 }, \qquad
\frac12 \pmatrix { 1 \cr 1 \cr -1 \cr 1 }
$$ 
as normalized eigenvectors which are maximally entangled.
Note that $\sigma_y \otimes \sigma_y$ also admits these
maximally entangled eigenvectors besides the Bell basis as
eigenvectors and the product eigenvectors.
\newline

For the triple spin interaction term 
$\sigma_x \otimes \sigma_y \otimes \sigma_z$ we obtain the eigenvalues
$+1$ (fourfold) and $-1$ (fourfold) and all the eight product states
as eigenstates given by the eigenstates of $\sigma_x$, $\sigma_y$,
$\sigma_z$. Owing to the degeneracies of the eigenvalues
we also find fully entangled states such as
$$
\frac12 \pmatrix { 1 & 1 & 0 & 0 & 0 & 0 & i & -i }^T 
$$
with the three-tangle as measure [11]. 

\section{Examples}

Consider now a specific example for $\alpha A$ and $\beta B$ with 
$n=2$ and $\epsilon > 0$.
Utilizing the Pauli spin matrices
$$
\alpha A = \hbar\omega_1 \sigma_z, \qquad
\beta B = \hbar\omega_2 \sigma_x 
$$
where $\alpha=\hbar \omega_1$, $\beta=\hbar \omega_2$ and 
$\omega_1$, $\omega_2$ are the frequencies. 
Note that $[\sigma_z,\sigma_x]=2i\sigma_y$ and
$\mbox{tr}(\hat H)=0$, $\mbox{tr}(\hat K)=0$. The elements of the set 
$$
\{ \, I_2 \otimes I_2, \sigma_z \otimes I_2, I_2 \otimes \sigma_x, \sigma_z \otimes \sigma_x \, \}
$$
form a commutative subgroup of the Pauli group ${\cal P}_2$. The elements 
$\sigma_z \otimes I_2$, $I_2 \otimes \sigma_x$, $\sigma_x \otimes \sigma_z$ are generators
of the Pauli group ${\cal P}_2$. Now the eigenvalues and eigenvectors of $\alpha A$ are given by
$$
\lambda_1 = \hbar \omega_1, \,\,\, {\bf u}_1 = \pmatrix { 1 \cr 0 }, \qquad
\lambda_2 = -\hbar \omega_1, \,\,\, {\bf u}_2 = \pmatrix { 0 \cr 1 }
$$
and the eigenvalues and eigenvectors of $\beta B$ are given by
$$
\mu_1 = \hbar \omega_2, \,\,\, {\bf u}_1 = \frac1{\sqrt2}\pmatrix { 1 \cr 1 }, 
\qquad \mu_2 = -\hbar \omega_2, \,\,\, 
{\bf u}_2 = \frac1{\sqrt2}\pmatrix { 1 \cr -1 } \,. 
$$
The Hamilton operator $\hat H$ is given by the $4 \times 4$ matrix 
which can be written as direct sum of two $2 \times 2$ matrices
$$
\widetilde H = 
\pmatrix { \hbar \omega_1 & \hbar \omega_2 + \epsilon & 0 & 0 \cr 
\hbar \omega_2 + \epsilon & \hbar \omega_1 & 0 & 0 \cr
0 & 0 & -\hbar \omega_1 & \hbar \omega_2 - \epsilon \cr 
0 & 0 & \hbar \omega_2 - \epsilon & -\hbar \omega_1 } \,.
$$ 
The eigenvalues of $\widetilde H$ are 
\begin{eqnarray*}
E_1(\omega_1,\omega_2,\epsilon) &=& \hbar \omega_1 + \hbar \omega_2 + \epsilon, \qquad
E_2(\omega_1,\omega_2,\epsilon) = \hbar \omega_1 - \hbar \omega_2 - \epsilon \\
E_3(\omega_1,\omega_2,\epsilon) &=& -\hbar \omega_1 + \hbar \omega_2 - \epsilon, \qquad
E_4(\omega_1,\omega_2,\epsilon) = -\hbar \omega_1 - \hbar \omega_2 + \epsilon
\end{eqnarray*}
with the corresponding eigenvectors (which can be written as product states)
$$
\pmatrix { 1 \cr 0 } \otimes \frac1{\sqrt2} \pmatrix { 1 \cr 1 }, \quad
\pmatrix { 1 \cr 0 } \otimes \frac1{\sqrt2} \pmatrix { 1 \cr -1 },
$$
$$
\pmatrix { 0 \cr 1 } \otimes \frac1{\sqrt2} \pmatrix { 1 \cr 1 }, \quad
\pmatrix { 0 \cr 1 } \otimes \frac1{\sqrt2} \pmatrix { 1 \cr -1 } \,.
$$
The Hamilton operator $\hat K$ is given by the $4 \times 4$ matrix
$$
\widetilde K = \pmatrix { \hbar \omega_1 & \hbar \omega_2 & \epsilon & 0 \cr 
\hbar \omega_2 & \hbar \omega_1 & 0 & -\epsilon \cr
\epsilon & 0 & -\hbar \omega_1 & \hbar \omega_2 \cr 
0 & -\epsilon & \hbar \omega_2 & -\hbar \omega_1 }
$$
with the four eigenvalues 
\begin{eqnarray*}
k_1(\omega_1,\omega_2,\epsilon) &=& -\sqrt{\hbar^2(\omega_1+\omega_2)^2 + \epsilon^2}, \\
k_2(\omega_1,\omega_2,\epsilon) &=& \sqrt{\hbar^2(\omega_1+\omega_2)^2 + \epsilon^2}, \\
k_3(\omega_1,\omega_2,\epsilon) &=& -\sqrt{\hbar^2(\omega_1-\omega_2)^2 + \epsilon^2}, \\
k_4(\omega_1,\omega_2,\epsilon) &=& \sqrt{\hbar^2(\omega_1-\omega_2)^2 + \epsilon^2} 
\end{eqnarray*}
and the corresponding unnormalized eigenvectors
$$
\pmatrix { \epsilon \cr \epsilon \cr k_1-\hbar(\omega_1+\omega_2) \cr 
k_2 + \hbar(\omega_1+\omega_2) }, \quad
\pmatrix { \epsilon \cr \epsilon \cr k_2-\hbar(\omega_1+\omega_2) \cr 
k_1+\hbar(\omega_1+\omega_2) },
$$
$$
\pmatrix { \epsilon \cr -\epsilon \cr k_3 - \hbar(\omega_1-\omega_2) \cr 
k_3 - \hbar(\omega_1-\omega_2) }, \quad
\pmatrix { \epsilon \cr -\epsilon \cr k_4-\hbar(\omega_1-\omega_2) \cr 
k_4-\hbar(\omega_1-\omega_2) } \,.
$$
Thus for the Hamilton operator $\hat H$ we have energy level 
crossing [10,12] which is due to the discrete symmetry of the 
Hamilton operator $\hat H$. For the Hamilton operator $\hat K$ 
we have no energy level crossing for $\epsilon > 0$. The symmetry 
is broken. For $\epsilon \to \infty$ and fixed frequencies the 
eigenvalues for the two Hamilton operators approach $\epsilon$ 
(twice) and $-\epsilon$ (twice). The four eigenvectors are entangled
for $\epsilon > 0$.
\newline

Extensions to higher order spin systems are straightforward.
An extension is to study the Hamilton operators
with triple spin interactions
\begin{eqnarray*}
\hat H &=& \hbar \omega_1 (\sigma_x \otimes I_2 \otimes I_2) 
+ \hbar \omega_2 (I_2 \otimes \sigma_y \otimes I_2) 
+ \hbar \omega_3 (I_2 \otimes I_2 \otimes \sigma_z) \\
&& + \gamma_{12}(\sigma_x \otimes \sigma_y \otimes I_2) 
+ \gamma_{13}(\sigma_x \otimes I_2 \otimes \sigma_z)
+ \gamma_{23}(I_2 \otimes \sigma_y \otimes \sigma_z) \\ 
&& + \epsilon (\sigma_x \otimes \sigma_y \otimes \sigma_z) 
\end{eqnarray*}
and
\begin{eqnarray*}
\hat K &=& \hbar \omega_1 (\sigma_x \otimes I_2 \otimes I_2) 
+ \hbar \omega_2 (I_2 \otimes \sigma_y \otimes I_2) 
+ \hbar \omega_3 (I_2 \otimes I_2 \otimes \sigma_z) \\
&& + \gamma_{12}(\sigma_x \otimes \sigma_y \otimes I_2) 
+ \gamma_{13}(\sigma_x \otimes I_2 \otimes \sigma_z)
+ \gamma_{23}(I_2 \otimes \sigma_y \otimes \sigma_z) \\ 
&& + \epsilon (\sigma_z \otimes \sigma_y \otimes \sigma_x) . 
\end{eqnarray*}
Triple spin interacting systems have been studied by several
authors [13,14,15]. For $\hat H$ we find the eight product
states given by the eigenstates of $\sigma_x$, $\sigma_y$,
$\sigma_z$. We also have energy level crossing owing to
the symmetry of the Hamilton operator $\hat H$. For the Hamilton
operator $\hat K$ the symmetry is broken and no level crossing
occurs. We also find entangled states for this Hamilton operator.
As an entanglement measure the three-tangle can be used [11]. 
Also the permutations $\sigma_z \otimes \sigma_x \otimes \sigma_y$,
$\sigma_y \otimes \sigma_z \otimes \sigma_x$ of the interacting
term could be investigated.
\newline

The question discussed in the introduction could also be studied 
for Bose systems with a Hamilton operator such as
$$
\hat H = \alpha (b^\dagger b \otimes I) + \beta (I \otimes (b^\dagger + b))
+ \gamma (b^\dagger b \otimes (b^\dagger + b)) 
$$
where $I$ is the identity operator and $\otimes$ denotes the tensor product.
\newline

In conclusion we have shown that swapping the terms in the
interacting part of Hamilton operators acting in a product
Hilbert space breaks the symmetry and thus the behaviour
about entanglement and energy level crossing will change.
 
\strut\hfill

{\bf References}

\strut \hfill

[1] E. Prugove\'cki, 
{\it Quantum Mechanics in Hilbert Space}, second edition,
Academic Press (1981)
\n
[2] W.-H. Steeb,
{\it Hilbert Spaces, Wavelets, Generalised Functions and Modern 
Quantum Mechanics}, Kluwer (1998)
\n
[3] W.-H. Steeb and Y. Hardy,
{\it Matrix Calculus and Kronecker Product}, second edition
Singapore, World Scientific (2011) 
\n 
[4] W.-H. Steeb, 
{\it Problems and Solutions in Introductory and Advanced Matrix Calculus},
 World Scientific, Singapore (2006) 
\n
[5] W.-H. Steeb and Y. Hardy, 
{\it Problems and Solutions in Quantum Computing and Quantum Information},
third edition, Singapore, World Scientific (2011)
\n
[6] W.-H. Steeb, {\it Problems and Solutions in Theoretical and 
Mathematical Physics}, third edition, Volume II: Advanced Level, 
Singapore, World Scientific (2009)
\n
[7] M. A. Nielsen and I. L. Chuang,
{\it Quantum Computing and Quantum Information}, 
Cambridge University Press (2000) 
\n
[8] D. Kielpinski,
J. Opt. B: Quantum Semiclass. Opt. {\bf 5} R121-R135 (2003)
\n
[9] N. D. Mermin,
Rev. Mod. Phys. {\bf 65}, 803-815 (1993)
\n
[10] W.-H. Steeb and Y. Hardy,
{\it Quantum Mechanics using Computer Algebra}, 
2nd edition, World Scientific, Singapore (2010)
\n
[11] A. Wong and N. Christensen, 
Phys. Rev. A 63, 044301 (2001)
\n
[12] W.-H. Steeb, A. J. van Tonder, C. M. Villet and S. J. M. Brits,
Found. Phys. Lett. {\bf 1}, 147-162 (1988)
\n
[13] C. Vanderzande and F. Igl\'oi,
J. Phys. A: Math. Gen. {\bf 20}, 4539-4549
\n
[14] R. Somma, G. Ortiz, E. Knill and J. Gubernatis,
Int. J. Quant. Inf. {\bf 1}, 189-206 (2003)
\n
[15] B. P. Lanyon, C. Hempel, D. Nigg, M. M\"uller, R. Gerritsma,
F. Z\"ahringer, P. Schindler, J. T. Barreiro, M. Rambach, 
G. Kirchmair, M. Heinrich, P. Zoller, R. Blatt and C. F. Roos,
arXiv:1109.1512v1 [quant-ph]
\n

\end{document}